\documentclass{aip-cp}

\usepackage[numbers]{natbib}
\usepackage{rotating}
\usepackage{graphicx}

\begin{document}

\title{Average and recommended half-life values for two-neutrino double beta decay: upgrade-2019}

\author[aff1]{A.S. Barabash\corref{cor1}}

\affil[aff1]{National Research Centre "Kurchatov Institute", Institute of Theoretical and Experimental Physics, B. Cheremushkinskaya 25, 117218 Moscow, Russia}
\corresp[cor1]{Corresponding author: barabash@itep.ru}

\maketitle

\begin{abstract}
All existing positive results on two neutrino double beta decay and two neutrino double electron capture in 
different nuclei were analyzed.  Using the procedure recommended by the 
Particle Data Group, weighted average values for half-lives of 
$^{48}$Ca, $^{76}$Ge, $^{82}$Se, $^{96}$Zr, $^{100}$Mo, $^{100}$Mo - 
$^{100}$Ru ($0^+_1$), $^{116}$Cd, $^{130}$Te, $^{136}$Xe, $^{150}$Nd, $^{150}$Nd - $^{150}$Sm 
($0^+_1$), $^{238}$U, $^{78}$Kr and $^{124}$Xe  were obtained. Existing geochemical data were 
analyzed and recommended values for half-lives of $^{128}$Te 
and $^{130}$Ba are proposed.  I recommend the use of these results as
the most currently reliable values for half-lives.
\end{abstract}

\section{INTRODUCTION}
In the present work, a critical analysis of all positive experimental
results has been performed, and averaged (or recommended) values for all 
isotopes are presented.

The first time this work was done was in 2001, and the 
results were presented at MEDEX'01 \cite{BAR02}. Then revised half-life values were
presented at MEDEX'05 \cite{BAR06}, MEDEX'09 \cite{BAR09a,BAR10} and MEDEX'13 \cite{BAR13,BAR15}. In the present paper, 
new positive results obtained since 2015 have been added and analyzed.  
\begin{table*}
\caption{Present, positive $2\nu\beta\beta$ decay results. 
Here, N is the number of useful events, S/B is the signal-to-background 
ratio. $^{a)}$ For $E_{2e} > 1.2$ MeV. $^{b)}$ After correction (see \cite{BAR15}). $^{c)}$ For $E_{2e} > 1.5$ MeV. 
$^{d)}$ For SSD mechanism. $^{e)}$ In both peaks. $^{f)}$ This value was obtained using average $T_{1/2}$ for $^{130}$Te and very well-known ratio $T_{1/2}$($^{130}$Te)/$T_{1/2}$($^{128}$Te) = $(3.52\pm 0.11)\cdot 10^{-4}$  \cite{BER93}.}
\bigskip
\label{Table1}
\begin{tabular}{|c|c|c|c|c|}
\hline
\rule[-2.5mm]{0mm}{6.5mm}
Nucleus & N & $T_{1/2}$, y & S/B & Ref., year \\
\hline
\rule[-2mm]{0mm}{6mm}
$^{48}$Ca & $\sim 100$ & $[4.3^{+2.4}_{-1.1}(stat)\pm 1.4(syst)]\cdot 10^{19}$
  & 1/5 & \cite{BAL96}, 1996 \\
 & 5 & $4.2^{+3.3}_{-1.3}\cdot 10^{19}$ & 5/0 & \cite{BRU00}, 2000 \\
& 153 & $[6.4^{+0.7}_{-0.6}(stat)^{+ 1.2}_{-0.9}(syst)\cdot 10^{19}$ & 3.9 & \cite{ARN16}, 2016 \\
\rule[-4mm]{0mm}{10mm}
 & & {\bf Average value:} $\bf 5.3^{+1.2}_{-0.8} \cdot 10^{19}$ & & \\  
          
\hline
\rule[-2mm]{0mm}{6mm}
$^{76}$Ge & $\sim 4000$ & $(0.9\pm 0.1)\cdot 10^{21}$ & $\sim 1/8$                                                        
& \cite{VAS90}, 1990 \\
& 758 & $1.1^{+0.6}_{-0.3}\cdot 10^{21}$ & $\sim 1/6$ & \cite{MIL91}, 1991 \\
& $\sim$ 330 & $0.92^{+0.07}_{-0.04}\cdot 10^{21}$ & $\sim 1.2$ & \cite{AVI91}, 1991 \\
& 132 & $1.27^{+0.21}_{-0.16}\cdot 10^{21}$ & $\sim 1.4$ & \cite{AVI94}, 1994 \\
& $\sim 3000$ & $(1.45\pm 0.15)\cdot 10^{21}$ & $\sim 1.5$ & \cite{MOR99}, 1999 
\\
& $\sim 80000$ & $[1.74\pm 0.01(stat)^{+0.18}_{-0.16}(syst)]\cdot 10^{21}$ & $\sim 1.5$ 
& \cite{HM03}, 2003 \\
& $\sim$ 30000 & $(1.925\pm 0.094)\cdot 10^{21}$ & $\sim 3$ & \cite{AGO15}, 2015 \\
\rule[-4mm]{0mm}{10mm}
& & {\bf Average value:} $\bf (1.88\pm 0.08) \cdot 10^{21}$ & & \\  
           
\hline
\rule[-2mm]{0mm}{6mm}
$^{82}$Se & 89.6 & $1.08^{+0.26}_{-0.06}\cdot 10^{20}$ & $\sim 8$ & \cite{ELL92}, 1992 \\
& 149.1 & $[0.83 \pm 0.10(stat) \pm 0.07(syst)]\cdot 10^{20}$ & 2.3 & 
\cite{ARN98}, 1998 \\
& 3472 & $[0.939 \pm 0.017(stat) \pm 0.058(syst)]\cdot 10^{20}$ & 4 & \cite{ARN18}, 2018\\ 
\rule[-4mm]{0mm}{10mm}
& & {\bf Average value:} $\bf (0.93\pm 0.05)\cdot 10^{20}$ & & \\
 
\hline
\rule[-2mm]{0mm}{6mm}
$^{96}$Zr & 26.7 & $[2.1^{+0.8}_{-0.4}(stat) \pm 0.2(syst)]\cdot 10^{19}$ & $1.9~^{a)}$ 
& \cite{ARN99}, 1999 \\
& 453 & $[2.35 \pm 0.14(stat) \pm 0.16(syst)]\cdot 10^{19}$ & 1 & \cite{ARG10}, 2010\\
\rule[-4mm]{0mm}{10mm}
& & {\bf Average value:} $\bf (2.3 \pm 0.2)\cdot 10^{19}$ & & \\

\hline
\rule[-2mm]{0mm}{6mm}
$^{100}$Mo & $\sim 500$ & $11.5^{+3.0}_{-2.0}\cdot 10^{18}$ & 1/7 & 
\cite{EJI91}, 1991 \\
& 67 & $11.6^{+3.4}_{-0.8}\cdot 10^{18}$ & 7 & \cite{ELL91}, 1991 \\
& 1433 & $[7.3 \pm 0.35(stat) \pm 0.8(syst)]\cdot 10^{18~b)}$ & 3 & 
\cite{DAS95}, 1995 \\
& 175 & $7.6^{+2.2}_{-1.4}\cdot 10^{18}$ & 1/2 & \cite{ALS97}, 1997 \\
& 377 & $[6.82^{+0.38}_{-0.53}(stat) \pm 0.68(syst)]\cdot 10^{18}$ & 10 & 
\cite{DES97}, 1997 \\
& 800 & $[7.2 \pm 1.1(stat) \pm 1.8(syst)]\cdot 10^{18}$ & 1/9 & 
\cite{ASH01}, 2001 \\
& $\sim$ 350 & $[7.15 \pm 0.37(stat) \pm 0.66(syst)]\cdot 10^{18}$ & 5~$^{c)}$ & 
\cite{CAR14}, 2014\\
& 9000 & $[6.90 \pm 0.15(stat) \pm 0.37(syst)]\cdot 10^{18~d)}$ & 10 & 
\cite{ARM17}, 2017\\
& 500000 & $[6.81 \pm 0.01(stat) ^{+0.38}_{-0.40}(syst)]\cdot 10^{18~d)}$ & 80 & 
\cite{ARN19}, 2019\\
\rule[-4mm]{0mm}{10mm}
& & {\bf Average value:} $\bf (6.88\pm 0.25)\cdot 10^{18}$ & & \\

\hline
\end{tabular}
\end{table*}

\addtocounter{table}{-1}
\begin{table*}
\caption{continued.}
\bigskip
\begin{tabular}{|c|c|c|c|c|}

\hline
$^{100}$Mo - & $133~^{e)}$ & $6.1^{+1.8}_{-1.1}\cdot 10^{20}$ & 1/7 & 
\cite{BAR95}, 1995 \\
$^{100}$Ru ($0^+_1$) &  $153~^{e)}$ & $[9.3^{+2.8}_{-1.7}(stat) \pm 1.4(syst)]\cdot 
10^{20}$ & 1/4 & \cite{BAR99}, 1999 \\
 & 19.5 & $[5.9^{+1.7}_{-1.1}(stat) \pm 0.6(syst)]\cdot 10^{20}$ & $\sim 8$ & 
\cite{DEB01}, 2001 \\ 
& 35.5 & $[5.5^{+1.2}_{-0.8}(stat) \pm 0.3(syst)]\cdot 10^{20}$ & $\sim 8$ & 
\cite{KID09}, 2009 \\ 
& 37.5 & $[5.7^{+1.3}_{-0.9}(stat) \pm 0.8(syst)]\cdot 10^{20}$ & $\sim 3$ & 
\cite{ARN07}, 2007 \\ 
& $597~^{e)}$ & $[6.9^{+1.0}_{-0.8}(stat) \pm 0.7(syst)]\cdot 10^{20}$ & $\sim 1/10$ & 
\cite{BEL10}, 2010 \\ 
& $239~^{e)}$ & $[7.5 \pm 0.6(stat) \pm 0.6(syst)]\cdot 10^{20}$ & $\sim 2$ & 
\cite{ARN14}, 2014 \\    
\rule[-4mm]{0mm}{10mm}
& & {\bf Average value:} $\bf 6.7^{+0.5}_{-0.4}\cdot 10^{20}$ & & \\

\hline
$^{116}$Cd& $\sim 180$ & $2.6^{+0.9}_{-0.5}\cdot 10^{19}$ & $\sim 1/4$ & 
\cite{EJI95}, 1995 \\
& 174.6 & $[2.9 \pm 0.3(stat) \pm 0.2(syst)]\cdot 10^{19~b)}$ & 3 & 
\cite{ARN96}, 1996 \\
& 9850 & $[2.9\pm 0.06(stat)^{+0.4}_{-0.3}(syst)]\cdot 10^{19}$ & $\sim 3$ & 
\cite{DAN03}, 2003 \\
& 4968 & $[2.74 \pm 0.04(stat) \pm 0.18(syst)]\cdot 10^{19~d)}$ & 12 & \cite{ARN17}, 2017\\
& 93000 & $(2.63^{+0.11}_{-0.12})\cdot 10^{19}$ & 1.5 & \cite{BAR18}, 2018\\
\rule[-4mm]{0mm}{10mm}
& & {\bf Average value:} $\bf (2.69 \pm 0.09)\cdot 10^{19}$ & & \\

\hline
\rule[-2mm]{0mm}{6mm}
$^{128}$Te & & $(2.41\pm 0.39)\cdot 10^{24}$ (geochem.)& & \cite{MES08}, 2008 \\
& & $(2.3\pm 0.3)\cdot 10^{24}$ (geochem.)& & \cite{THO08}, 2008 \\
\rule[-4mm]{0mm}{10mm}
& & {\bf Recommended value:} $\bf (2.25\pm 0.09)\cdot 10^{24~f)}$ & & \\

\hline
\rule[-2mm]{0mm}{6mm}
$^{130}$Te& 260 & $[6.1 \pm 1.4(stat)^{+2.9}_{-3.5}(syst)]\cdot 10^{20}$ & 1/8 & \cite{ARN03}, 2003 \\
& 236 & $[7.0 \pm 0.9(stat) \pm 1.1(syst)]\cdot 10^{20}$ & 1/3 & \cite{ARN11}, 2011 \\
& $\sim$ 33000 & $[8.2 \pm 0.2(stat) \pm 0.6(syst)]\cdot 10^{20}$ & 0.1-0.3 & \cite{ALD17}, 2017 \\
& $\sim$ 20000 & $[7.9 \pm 0.1(stat) \pm 0.2(syst)]\cdot 10^{20}$ & $>$1 & \cite{CAM19}, 
2019 \\
\rule[-4mm]{0mm}{10mm}
& & {\bf Average value:} $\bf (7.91\pm 0.21)\cdot 10^{20}$ & & \\

\hline
\rule[-2mm]{0mm}{6mm}
& $\sim$ 19000 & $[2.165 \pm 0.016(stat) \pm 0.059(syst)]\cdot 10^{21}$ & 10 & 
\cite{ALB13}, 2014 \\
$^{136}$Xe & $\sim$ 100000 & $[2.21 \pm 0.02(stat) \pm 0.07(syst)]\cdot 10^{21}$ & 
10 & \cite{GAN12}, 2016 \\
\rule[-4mm]{0mm}{10mm}
& & {\bf Average value:} $\bf(2.18\pm 0.05)\cdot 10^{21}$ & & \\

\hline
\rule[-2mm]{0mm}{6mm}
$^{150}$Nd& 23 & $[18.8^{+6.9}_{-3.9}(stat) \pm 1.9(syst)]\cdot 10^{18}$ & 
1.8 & \cite{ART95}, 1995 \\
& 414 & $[6.75^{+0.37}_{-0.42}(stat) \pm 0.68(syst)]\cdot 10^{18}$ & 6 & 
\cite{DES97}, 1997 \\
& 2214 & $[9.34 \pm 0.22(stat)^{+0.62}_{-0.60}(stat) \pm 0.63(syst)]\cdot 10^{18}$ & 4 & \cite{ARN16a}, 2016\\
\rule[-4mm]{0mm}{10mm}
& & {\bf Average value:} $\bf(8.4\pm 1.1)\cdot 10^{18}$ & & \\
& & {\bf Recommended value:} $\bf(9.34^{+0.67}_{-0.64}) \cdot 10^{18}$ & & \\

\hline
\rule[-2mm]{0mm}{6mm}
$^{150}$Nd - & $177.5~^{e)}$ & $[1.33^{+0.36}_{-0.23}(stat)^{+0.27}_{-0.13}(syst)]\cdot 10^{20}$ & 1/5 & \cite{BAR09}, 2009 \\
& 21.6 & $[1.07^{+0.45}_{-0.25}(stat) \pm 0.07(syst)]\cdot 10^{20}$ & 1.2 & \cite{KID14}, 2014 \\
$^{150}$Sm ($0^+_1$) & & {\bf Average value:} $\bf 1.2^{+0.3}_{-0.2}\cdot 10^{20}$ & \\ 
 
\hline
\rule[-2mm]{0mm}{6mm}
$^{238}$U& & $\bf (2.0 \pm 0.6)\cdot 10^{21}$ (radiochem.) & & \cite{TUR91}, 1991 \\

\hline
\end{tabular}
\end{table*}

\begin{table}
\caption{Present, positive two neutrino double electron capture  results. 
Here, N is the number of useful events, S/B is the signal-to-background 
ratio. In case of $^{78}$Kr and $^{124}$Xe $T_{1/2}$ for $2K(2\nu)$ capture is presented (this is $\sim$ 75-80\% of $ECEC(2\nu)$). $^a)$ See text.}
\bigskip
\label{Table2}
\begin{tabular}{|c|c|c|c|c|}

\hline
\rule[-2mm]{0mm}{6mm}
$^{130}$Ba &  & $\bf 2.1^{+3.0}_{-0.8} \cdot 10^{21}$ (geochem.) & 
 & \cite{BAR96}, 1996 \\
 $ECEC(2\nu)$&  & $\bf (2.2 \pm 0.5)\cdot 10^{21}$ (geochem.) & 
 & \cite{MES01}, 2001 \\
& & $\bf (0.60 \pm 0.11)\cdot 10^{21}$ (geochem.) &
& \cite{PUJ09}, 2009 \\
\rule[-4mm]{0mm}{10mm}
& & {\bf Recommended value:} $\bf (2.2 \pm 0.5)\cdot 10^{21}$  & & \\

\hline
\rule[-2mm]{0mm}{6mm}
$^{78}$Kr & 15 & $\bf [1.9^{+1.3}_{-0.7}(stat \pm 0.3(syst)] \cdot 10^{22}$ &  
 & \cite{RAT17}, 2017 \\
$2K(2\nu)$ & & & & \\
\rule[-4mm]{0mm}{10mm}
& & {\bf Recommended value:} $\bf (1.9^{+1.3}_{-0.8})\cdot 10^{22}$ (?)$^a)$ & 15 & \\

\hline
\rule[-2mm]{0mm}{6mm}
$^{124}$Xe & 126 & $\bf [1.8 \pm 0.5(stat \pm 0.1(syst)] \cdot 10^{22}$ & 0.2 
 & \cite{APR19}, 2019 \\
$2K(2\nu)$ & & & & \\
\rule[-4mm]{0mm}{10mm}
& & {\bf Recommended value:} $\bf (1.8 \pm 0.5)\cdot 10^{22}$  & & \\

\hline
\end{tabular}
\end{table}

\section{PRESENT EXPERIMENTAL DATA}
Experimental results on $2\nu\beta\beta$ decay and $ECEC(2\nu)$ in different nuclei are 
presented in Table 1 and Table 2, respectively.  For direct experiments, the number of useful events and the signal-to-background ratio are presented. The results of geochemical experiments for $^{82}$Se, $^{96}$Zr, $^{100}$Mo and $^{130}$Te are not presented in Table 1, since these data were not used in the calculation of average values (see the discussion in \cite{BAR10,BAR15}).

\section{DATA ANALYSIS}
To obtain an average of the ensemble of available data, a standard weighted 
least-squares procedure, as recommended by the Particle Data Group 
\cite{PDG18}, was used (see also \cite{BAR15}). In case of $^{76}$Ge  only results from  \cite{HM03,AGO15} (as most precise and reliable) were used to obtain the average half-life value. In case of  $^{100}$Mo I used results from \cite{DES97,CAR14,ARM17,ARN19} only for the same reason. Due to the wide variation in results for $^{150}$Nd, I recommend using the result obtained in the NEMO-3 experiment as the most reliable estimate of the half-life of $^{150}$Nd
by this moment. 

It has to be stressed that for $^{238}$U, $^{124}$Xe and $^{78}$Kr "positive" result was obtained in the only experiment. This is why confirmation of these results in independent experiments is needed. And one has to be very careful using indicated half-life values. Most questionable situation is for $^{78}$Kr. Number of detected events is quite small and analysis of experimental data is very complicated. If obtained result is correct it means that in case of $^{78}$Kr we have a deal with highest value of Nuclear Matrix Element (NME) among all nuclei for which 2$\nu$ decay was observed (NME for $^{78}$Kr  is $\sim$ 2 times higher than in case of $^{100}$Mo, which has the highest value of NME). In principle it is possible, but looks a little bit strange.  

\section{ACKNOWLEDGMENTS}
This work was partly supported by Russian Scientific Foundation (grant No. 18-12-00003)

\nocite{*}
\bibliographystyle{aipnum-cp}%
\bibliography{sample}%

\end{document}